\documentclass[aps,pre,reprint,superscriptaddress,longbibliography]{revtex4-2}
\usepackage{amsmath,amssymb,amsfonts}
\usepackage{graphicx}
\usepackage{bm}
\usepackage{color}
\usepackage{hyperref}

\begin{document}

\title{Borel Summability of the Spatial Gradient Expansion in Linear Kinetic Theory}
\author{Mahdi Kooshkbaghi}
\email{mahdi.kooshkbaghi@gmail.com}
\affiliation{Independent Researcher, Princeton, New Jersey 08540, USA}
\begin{abstract}
    The passage from kinetic theory to hydrodynamics proceeds through
    gradient expansions whose large-order behavior depends on the model and on
    whether the gradients are temporal or spatial.
    Known examples of divergent temporal gradient expansions include those arising 
    in Bjorken flow in M\"uller--Israel--Stewart theory, where positive-axis Borel singularities obstruct ordinary Borel summation and require a transseries completion.
    By contrast, some spatial gradient
    expansions in holographic theories and relativistic kinetic theory converge
    within a finite radius.
    We show that the spatial Chapman--Enskog expansion of the linear
    one-dimensional BGK equation realizes a third possibility: its coefficients
    can be obtained in closed form at all orders, and the resulting series is
    factorially divergent yet Borel summable along the positive axis.
    The unambiguous Borel sum reconstructs the hydrodynamic dispersion branch
    up to its merger with the essential spectrum and provides its analytic
    continuation beyond that point.
    We trace the divergence of the spatial expansion to the unbounded velocity
    support of the Maxwellian equilibrium: replacing it with a compactly
    supported equilibrium removes the factorial growth and gives the series a
    nonzero radius of convergence.
\end{abstract}
\maketitle

\section{Introduction}
The classical treatment of non-equilibrium systems in kinetic theory relies
on a systematic gradient expansion of the distribution function,
known as the Chapman--Enskog (CE) expansion~\cite{chapman1990mathematical},
truncated at finite order to yield the macroscopic equations of fluid motion.
The validity of this expansion is controlled by the Knudsen number, the ratio
of microscopic kinetic scales to the macroscopic scales of the coarse-grained
hydrodynamic description. In relativistic systems, the corresponding expansion
is organized in dimensionless gradients formed using the local temperature as
the microscopic scale.
However, it is well established that the resulting gradient expansion can be
asymptotic and divergent, both in classical kinetic
theory~\cite{santos1986} and for a massless relativistic gas~\cite{denicol2016divergence}.

Nevertheless, numerical studies show that hydrodynamics can become
quantitatively accurate on time scales controlled by the decay of
non-hydrodynamic degrees of freedom~\cite{Chesler:2010bi}, while experimental
flow measurements provide evidence for collective behavior in high-energy
collisions~\cite{aad2016observation}.
Together with decades of work on higher-order macroscopic closures in classical
kinetic theory~\cite{struchtrup2005, gorban2005invariant}, these results motivate reduced
descriptions beyond the leading gradient regime.
The coexistence of successful hydrodynamic behavior with divergent gradient
series motivates the question of summability: whether an appropriate summation
procedure can recover the hydrodynamic response from such a series.

This question was first explored for \emph{temporal} gradient expansions by
Heller and Spali\'{n}ski~\cite{heller2015hydrodynamics}.  In the
M\"uller--Israel--Stewart framework, the symmetries of Bjorken flow reduce the
fluid dynamics to a set of ordinary differential equations.  The corresponding
gradient series has
positive-axis Borel singularities associated with decaying non-hydrodynamic
modes.  The ordinary Borel sum is therefore ambiguous in the physical
direction, and a transseries completion is required~\cite{basar2015hydrodynamics}, which ultimately yields
the hydrodynamic attractor; see Ref.~\cite{aniceto2019primer} for a review of resurgent transseries and their asymptotics.  A similar structure was
subsequently found in kinetic theory within the relaxation-time
approximation~\cite{heller2018hydrodynamization}.

The hydrodynamic attractor was further examined by Romatschke~\cite{romatschke2018fluid}
for conformal Bjorken flow in three theoretical frameworks: 
resummed BRSSS (rBRSSS) hydrodynamics~\cite{Baier:2007ix}, kinetic theory in the
relaxation-time approximation, and strongly coupled $\mathcal{N}=4$ supersymmetric
Yang--Mills theory via AdS/CFT. 
The divergence of temporal gradient expansions
has since been shown to persist in less symmetric settings and is therefore not
merely an artifact of Bjorken flow~\cite{heller2022diverges}.

The \emph{spatial} (wavenumber) gradient expansion can behave differently.
Grozdanov \textit{et al.} analyzed hydrodynamic dispersion relations at complex frequency
and momentum and showed that, in holographic theories, their gradient expansions can possess
a finite, nonzero radius of convergence~\cite{grozdanov2019convergence,grozdanov2019complex}.
Heller \textit{et al.} subsequently showed that, in linear response, convergence is controlled
by a theory-dependent critical momentum and demonstrated this finite-radius structure in
relativistic kinetic theory~\cite{heller2021linear, heller2021convergence}.
The non-relativistic setting is in some respects more subtle. Certain higher-order truncations of the spatial expansion are unstable at short wavelengths~\cite{bobylev1982,bobylev2006instabilities}, which has long motivated non-perturbative approaches.

For the one-dimensional BGK kinetic equation, we derived in earlier work with
Karlin and Chikatamarla~\cite{karlin2014non} the exact invariance equation for the
hydrodynamic manifold and constructed its solution through an iterative numerical
procedure.
A complementary non-perturbative approach was developed by Kogelbauer and
Karlin~\cite{kogelbauer2021nonlocal,kogelbauer2024rigorous} in the framework of spectral
theory: the hydrodynamic manifold was identified as the slow eigenmode of
the associated Jacobi operator, yielding the exact diffusion eigenvalue
and a critical wavenumber $k_{\rm crit}$ beyond which the slow mode ceases to
exist as an isolated spectral mode.
More recently, the CE series was shown~\cite{kogelbauer2025relation} to agree
formally, order by order in the Knudsen number, with this spectral closure,
while diverging factorially for every nonzero Knudsen number.

Although the terminology used for temporal and spatial gradient expansions
differs, both settings implement a common
\emph{model-reduction principle}. Fast non-hydrodynamic degrees of freedom are
eliminated to obtain effective dynamics for slow hydrodynamic variables.
For temporal gradient expansions, particularly in Bjorken flow, a
\emph{hydrodynamic attractor} is a distinguished solution that other solutions
approach in forward time as non-hydrodynamic transients decay. A suitable
non-perturbative summation of the temporal gradient expansion can reconstruct
this attractor~\cite{heller2015hydrodynamics,heller2018hydrodynamization,romatschke2018fluid}.

In classical kinetic theory, spatial gradient expansions are instead naturally
viewed as perturbative representations of a \emph{slow invariant manifold}%
~\cite{gorban2005invariant,karlin2014non}. In linear response, the slow sector
is characterized by isolated modes and their dispersion relations. A closure
expresses the non-conserved moments in terms of the conserved fields. Together
with the conservation laws, it determines the hydrodynamic dispersion
relations. When the slow modes are spectrally separated from the remaining
kinetic modes, the latter decay more rapidly toward the slow subspace%
~\cite{kogelbauer2021nonlocal,kogelbauer2024rigorous}.
The temporal-attractor and spatial-manifold viewpoints are, therefore, two versions of the same model-reduction idea. But the attractor, invariant manifold, closure, and dispersion relation are still distinct mathematical objects.

The present work revisits the linear one-dimensional BGK model studied in
Ref.~\cite{karlin2014non}. The model is isothermal, has a constant relaxation
time, and is linearized about a homogeneous equilibrium with zero mean
velocity. Its relaxation target is chosen to conserve only the number density. Rather than
studying attractive temporal dynamics, we develop an analytic treatment of
its spatial Chapman--Enskog expansion and Borel resummation. Specifically, we
(i) introduce a polynomial-recurrence construction for the invariance equation,
replacing the iterative numerical procedure of
Ref.~\cite{karlin2014non};
(ii) obtain the exact CE coefficients at all orders by Lagrange inversion;
(iii) derive their large-order behavior;
(iv) establish that the resummed CE expansion coincides with the spectral
hydrodynamic branch throughout its domain and continues analytically beyond
its critical endpoint; and
(v) trace the factorial divergence to the unbounded velocity support of the
Maxwellian and show that compact velocity support gives the CE series a
nonzero radius of convergence, as illustrated by a restricted
one-dimensional, isothermal bounded-velocity BGK analogue.

\section{Invariant-Manifold Formulation and Polynomial-Recurrence Construction}
\label{sec:manifold}
Equations~\eqref{eq:BGK_nondim}--\eqref{eq:Omega_IC}, the generating-function formulation and the exact invariance ODE, follow Ref.~\cite{karlin2014non} and are restated to fix notation.
Section~\ref{sec:recurrence} departs from that work: a polynomial-recurrence closure of the invariance equation generates a sequence of truncation polynomials $\{P_n\}$ that determines the physical branch algebraically, in place of the iterative ``pullout'' continuation employed previously.

\subsection{Kinetic equation and generating-function formulation}
Starting from the one-dimensional isothermal BGK~\cite{bgk1954model} equation with constant temperature $T$ and relaxation time $\tau$, and choosing units such that $m=k_B T=\tau=1$, we write the non-dimensionalized kinetic equation for the distribution function $f(x,v,t)$ as
\begin{equation}\label{eq:BGK_nondim}
  \partial_t f = -v\partial_x f - (f - f^{\rm eq}),
\end{equation}
where the local equilibrium is the Maxwellian $f^{\rm eq} = n(x,t)\,(2\pi)^{-1/2}\,e^{-v^2/2}$, and $n(x,t) = \int_{-\infty}^{\infty} f\,dv$ is the locally conserved particle density. 
Taking the velocity moments of the distribution function,
\begin{equation*}
M_l(x,t) = \int_{-\infty}^{\infty} v^l f\,dv, \quad l=0,1,2,\ldots,
\end{equation*}
in~\eqref{eq:BGK_nondim} yields an infinite,
unclosed hierarchy of moment equations, whose equilibrium moments take the Gaussian form,
\begin{equation*}
  M_l^{\rm eq}(x,t) =
  \begin{cases}
    n(x,t)(l-1)!! & l \text{ even}, \\
    0             & l \text{ odd}
  \end{cases}.
\end{equation*}
Following the machinery typically used in the \emph{method of invariant manifolds}~\cite{gorban2005invariant}, the infinite hierarchy is compactly encoded in the generating function
$Z(\lambda, x, t) = \int_{-\infty}^{\infty} e^{-i\lambda v} f\,dv$. Applying the spatial Fourier transform to the even (real) and odd (imaginary) parts of $Z$ parametrizes the hydrodynamic manifold so that all moments depend on the Fourier-transformed conserved density~$\hat{n}$,
\begin{align}
  \partial_t \hat{Z}_{\text{re}} & = ik\,\partial_\lambda \hat{Z}_{\text{im}} - \hat{Z}_{\text{re}} + \hat{n}\,e^{-\lambda^2/2},
  \label{eq:Zhat_plus}   \\
  \partial_t \hat{Z}_{\text{im}} & = -ik\,\partial_\lambda \hat{Z}_{\text{re}} - \hat{Z}_{\text{im}}.
  \label{eq:Zhat_minus}
\end{align}
In Fourier space this means
$\hat{Z}_{\text{re}} = \hat{\Theta}_{\text{re}}(\lambda, k^2)\,\hat{n}$ and
$\hat{Z}_{\text{im}} = ik\,\hat{\Theta}_{\text{im}}(\lambda, k^2)\,\hat{n}$,
where the factor $ik$ in $\hat{Z}_{\text{im}}$ reflects the odd parity of
the flux.
The core requirement is the \emph{dynamic invariance condition}: the kinetic vector field must be tangent to this manifold, equating the microscopic time derivative with that induced by the reduced
density dynamics~\cite{Karlin2002}. Therefore, summation of the Chapman--Enskog expansion
and direct solution of the invariance equation are two routes to derive hydrodynamic from kinetics, the latter being geometrically
motivated (see Ref.~\cite{gorban2014hilbert} for the
connection to Hilbert's sixth problem.)

\subsection{Invariance equation and the pullout procedure}
On the hydrodynamic manifold, we define $\hat{\omega}(k)$ by the density evolution
$\hat{n}(k,t)=\hat{n}(k,0)e^{\hat{\omega}(k)t}$.
Imposing the invariance condition yields a system of two first-order ODEs in $\lambda$ for
$\hat{\Theta}_{\text{re,im}}$. 
Eliminating $\hat{\Theta}_{\text{re}}$ and
introducing the frequency function $\hat{\Omega}(\lambda, k^2)$ via
$\partial_\lambda\hat{\Theta}_{\text{im}}= -k^{-2}\,\hat{\Omega}(\lambda,k^2)\,e^{-\lambda^2/2}$,
one obtains a single exact ODE for $\hat{\Omega}$:
\begin{equation}\label{eq:Omega_ODE_eq}
  (\hat{\omega} + 1)^2\hat{\Omega}
  + k^2(1 - \lambda^2)(\hat{\Omega} + 1)
  = k^2\left(\partial_\lambda^2\hat{\Omega}
  - 2\lambda\partial_\lambda\hat{\Omega}\right),
\end{equation}
subject to the initial conditions
\begin{equation}\label{eq:Omega_IC}
  \hat{\Omega}(0, k^2) = \hat{\omega}(k), \qquad
  \left.\partial_\lambda\hat{\Omega}\right|_{\lambda=0} = 0.
\end{equation}

Equation~\eqref{eq:Omega_ODE_eq} is the exact, non-perturbative hydrodynamic equation for this system; however, the initial conditions~\eqref{eq:Omega_IC} do not fix $\hat{\omega}(k)$ uniquely, and in prior work the physical branch was isolated numerically by an iterative ``pullout'' procedure~\cite{karlin2014non}: successive Taylor coefficients of $\hat{\Omega}$ in $\lambda$ are set to zero, and the root with $\hat{\omega}\to 0$ as $k\to 0$ is tracked in $k$ by numerical continuation~\cite{doedel2007auto}, each order pulling the branch out to larger wavenumbers until it collides with a kinetic root.

\subsection{Polynomial-recurrence closure}
\label{sec:recurrence}
We now show that this phase-space dimensionality reduction can be carried out entirely algebraically, without recourse to numerical continuation.
Since the frequency function is even and analytic in $\lambda$, we can expand 
it in a power series
\begin{equation}\label{eq:Omega_series}
  \hat{\Omega}(\lambda, k^2) = \hat{\omega}
  + \sum_{n=1}^{\infty} \frac{\lambda^{2n}\,\hat{\omega}_{2n}(\hat{\omega}, k^2)}
  {(2n)!\,k^{2n}},
\end{equation}
Substituting~Eq.~\eqref{eq:Omega_series} into Eq.~\eqref{eq:Omega_ODE_eq}, one finds that the coefficients factorize as $\hat{\omega}_{2n} = (\hat{\omega}+1)P_n(\hat{\omega}, k^2)$, where $\{P_n\}_{n\geq 0}$ is a sequence of truncation polynomials generated by the three-term recurrence ($n \geq 2$):
\begin{equation}
  \label{eq:poly}
  P_n = \bigl[(\hat{\omega}+1)^2 + (4n{-}3)k^2\bigr]P_{n-1} - k^4(2n{-}2)(2n{-}3)P_{n-2},
\end{equation}
with $P_0 = 1$ and $P_1 = \hat{\omega}(\hat{\omega}+1) + k^2$.
The three-term structure is a direct consequence of the Hermite-type
differential operator $\partial_\lambda^2 - 2\lambda\partial_\lambda$
in Eq.~\eqref{eq:Omega_ODE_eq}.
The common root $\hat{\omega}=-1$ of the truncation conditions coincides
with the relaxation eigenvalue of the collision operator. For $k\ne0$,
the full kinetic operator has essential spectrum on
$\operatorname{Re}\hat{\omega}=-1$~\cite{kogelbauer2025relation}.

\begin{figure}[!htp]
  \centering
  \includegraphics[width=\linewidth]{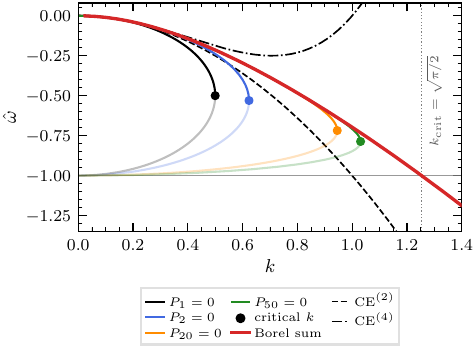}
  \caption{Hydrodynamic dispersion relation $\hat{\omega}(k)$ of the linear
  one-dimensional BGK equation. The closed-form Borel sum from
  Eq.~\eqref{eq:omega_closed} (red) coincides with the spectral hydrodynamic branch for
  $0<k<k_{\rm crit}=\sqrt{\pi/2}$. At $k_{\rm crit}$, the mode merges
  with the essential spectrum at $\hat{\omega}=-1$ (grey); beyond this
  point, the red curve represents the analytic continuation of the dispersion
  branch and no longer corresponds to an isolated hydrodynamic mode.
  Solid coloured curves show the physical roots of the spectral
  polynomial branches Eq.~\eqref{eq:poly} $P_n(\hat{\omega},k^2)=0$ for $n=1,2,20,50$, converging
  to the Borel sum with increasing truncation order; faded segments past the
  fold bifurcation at $k_{\rm turn}(n)$ (filled circles) indicate the
  unphysical continuation, with $k_{\rm turn} \approx 0.50,\,0.62,\,0.95,\,1.03$
  for $n=1,2,20,50$ respectively. Finite-order Chapman--Enskog truncations
  $\mathrm{CE}^{(2)}\!:\hat{\omega}=-k^2$ (dashed, classical diffusion) and
  $\mathrm{CE}^{(4)}\!:\hat{\omega}=-k^2+k^4$ (dash-dotted, Burnett-type) both
  diverge from the branch at moderate $k$, illustrating the factorial
  divergence of the CE series that Borel summation resolves.}
  \label{fig:attractor}
\end{figure}

Truncating the hierarchy at order $n$ by setting $\hat{\omega}_{2n} = 0$ yields the algebraic equation $P_n(\hat{\omega}, k^2) = 0$.
Here, we explore this polynomial systematically via numerical continuation in $k$ using \texttt{Auto-07p}~\cite{doedel2007auto}, following the physical root $\hat{\omega}(k)$ seeded by $\hat{\omega} = 0$ at $k = 0$.
As shown in Fig.~\ref{fig:attractor}, each branch terminates at a turning point $k_\mathrm{turn}(n)$ that grows with $n$, and successive truncations agree over a widening range of $k$.

\section{Exact Chapman--Enskog Coefficients}
\label{sec:ce}
\subsection{Reduction to a constant-coefficient equation}
The differential operator
$\mathcal{L}=\partial_\lambda^2 - 2\lambda\,\partial_\lambda$ on the
right-hand side of Eq.~\eqref{eq:Omega_ODE_eq} satisfies the
conjugation identity
$\mathcal{L}[e^{\lambda^2/2}\Phi]
= e^{\lambda^2/2}[\Phi''+(1-\lambda^2)\Phi]$.
Since the factor $(1-\lambda^2)$ matches the potential multiplying
$(\hat{\Omega}+1)$, we introduce the substitution
\begin{equation*}
  \hat{\Omega}(\lambda)
  = (\hat{\omega}+1)\,e^{\lambda^2/2}\,\Phi(\lambda) - 1
\end{equation*}
which eliminates all $\lambda$-dependent coefficients, reducing
Eq.~\eqref{eq:Omega_ODE_eq} to the constant-coefficient ODE
\begin{equation}\label{eq:Phi_ODE}
  \Phi''(\lambda) - A\Phi(\lambda)
  = -\frac{A}{\hat{\omega}+1}e^{-\lambda^2/2},
  \quad A = \frac{(\hat{\omega}+1)^2}{k^2},
\end{equation}
with $\Phi(0)=1$ and $\Phi'(0)=0$.

The \textit{homogeneous} solutions of Eq.~\eqref{eq:Phi_ODE} are
$e^{\pm\sqrt{A}\,\lambda}$. Since $f(x,v,t)$ is
integrable in $v$, the Riemann--Lebesgue lemma requires $Z(\lambda)$
to remain bounded as $|\lambda|\to\infty$; $\hat{\Omega}$ may grow
at most as $e^{\lambda^2/2}$, and consequently $\Phi(\lambda)$ must be bounded.
This \textit{rules out both homogeneous solutions}.

\subsection{Exact self-consistency relation}
The unique bounded \textit{particular solution} is obtained by convolving the
source with the free-space Green's function
$G(\lambda,s) = -\frac{1}{2\sqrt{A}}\,e^{-\sqrt{A}\,|\lambda - s|}$
of the operator $d^2/d\lambda^2 - A$:
\begin{equation}\label{eq:Phi_integral}
  \Phi(\lambda) = \frac{\sqrt{A}}{2(\hat{\omega}+1)}
  \int_{-\infty}^{\infty}
  e^{-\sqrt{A}\,|\lambda-s|}\;
  e^{-s^2/2}\,ds.
\end{equation}
Evaluating Eq.~\eqref{eq:Phi_integral} at $\lambda=0$ and imposing
$\Phi(0)=1$, the two-sided exponential kernel is recast as a
velocity-space resolvent via its Fourier representation
$\tfrac{\sqrt{A}}{2}\,e^{-\sqrt{A}|s|}
= \frac{1}{2\pi}\!\int\!\frac{A}{A+v^2}\,e^{ivs}\,dv$,
followed by Gaussian $s$-integration.
This yields the exact self-consistency condition
\begin{equation*}
  \hat{\omega}+1 = \int_{-\infty}^{\infty}
  \frac{A}{A+v^2}\;\frac{e^{-v^2/2}}{\sqrt{2\pi}}\,dv,
\end{equation*}
For small $k$ (equivalently large~$A$), the resolvent is expanded as
a geometric series,
$\frac{A}{A+v^2}=\sum_{m=0}^{\infty}(-1)^m(v^2/A)^m$, and
integrated term by term using the Gaussian moments
$\langle v^{2m}\rangle=(2m{-}1)!!$\,:
\begin{align}\label{eq:self_consistency}
  \hat{\omega}
  &= \sum_{m=1}^{\infty}(-1)^m\,(2m{-}1)!!\;
  \frac{k^{2m}}{(1{+}\hat{\omega})^{2m}} \notag\\
  &= \sum_{m=1}^{\infty}(-1)^m\,(2m{-}1)!!\;x^m
  \;:=\; F(x),
\end{align}
where $x = k^2(1{+}\hat{\omega})^{-2}=A^{-1}$. 
Since $x$ itself depends on $\hat{\omega}$, Eq.~\eqref{eq:self_consistency} is an implicit relation for $\hat{\omega}(k)$.
\subsection{All-order coefficients by Lagrange inversion}
The Chapman--Enskog coefficients
$\hat{\omega}=\sum_{n=1}^{\infty}a_{2n}\,k^{2n}$ are extracted by
Lagrange inversion:
\begin{equation}\label{eq:a2n_exact}
  a_{2n} = \frac{1}{n}\,[x^{n-1}]\bigl\{
  F'(x)\,(1{+}F(x))^{-2n}\bigr\},
\end{equation}
where $[x^{m}]\,g(x)$ denotes the coefficient of $x^{m}$ in the
power series of~$g$
(see Chap.~5 of Ref.~\cite{wilf2005generatingfunctionology} for details on coefficient extraction notation and Lagrange inversion of power series). 

Equation~\eqref{eq:a2n_exact} provides, to our knowledge, novel closed-form expressions for the exact CE transport coefficients at all orders, determined entirely by the equilibrium velocity moments $(2m{-}1)!!$. The explicit evaluation of the first few coefficients yields
$a_2 = -1, a_4 = 1, a_6 = -4, a_8 = 27, a_{10} = -248,
a_{12} = 2830, a_{14} = -38232$, matching the sequence
reported in Ref.~\cite{karlin2014non} and recently confirmed by spectral study in~\cite{kogelbauer2025relation}.

\begin{figure}[!htp]
  \centering
  \includegraphics[width=\linewidth]{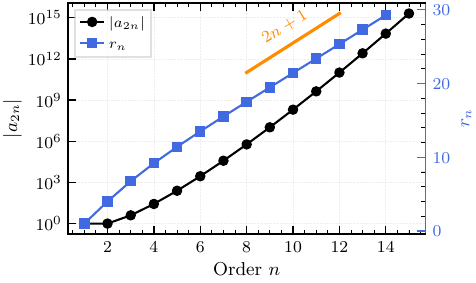}
  \caption{Absolute values of the exact CE coefficients $|a_{2n}|$ from Eq.~\eqref{eq:a2n_exact} (black circles, left axis) and ratios $r_n = |a_{2(n+1)}/a_{2n}|$ (blue squares, right axis). The convex shape of $|a_{2n}|$ on a log scale signals factorial divergence, while the asymptotic growth $r_n\sim2n+1$ implies zero radius of convergence.
  This factorial scaling is the spatial analog of the temporal divergence in Fig.~2 of Ref.~\cite{heller2015hydrodynamics}; unlike that case, the present series is Borel summable along the positive axis (Sec.~\ref{sec:borel}).}
  \label{fig:ce_coeffs}
\end{figure}

As shown in Fig.~\ref{fig:ce_coeffs}, the coefficients grow factorially and
the ratio $r_n=|a_{2(n+1)}/a_{2n}|$ increases roughly linearly; the asymptotic
law $r_n\sim2n+1$, established in Sec.~\ref{sec:largeorder}, implies a zero
radius of convergence for the spatial CE series.

\section{Borel Summation}
\label{sec:borel}

\subsection{Borel transform of the moment series}
Gaussian moments $(2m-1)!!$ are responsible for the factorial growth of the CE coefficients.
The Borel transform compensates for this
growth by dividing the coefficient of $x^m$ by $m!$.  Introducing the Borel
variable $\sigma$, we define
\begin{equation}
  \mathcal{B}[F](\sigma)
  := \sum_{m=1}^{\infty}
  \frac{(-1)^m(2m{-}1)!!}{m!}\,\sigma^m.
\end{equation}
Using $(2m{-}1)!!/m!=\binom{2m}{m}/2^m$, we obtain
\begin{align}
  \label{eq:Borel_F}
  \mathcal{B}[F](\sigma)
  &= \sum_{m=1}^{\infty}\binom{2m}{m}
  \left(-\frac{\sigma}{2}\right)^m \\
  &= \sum_{m=0}^{\infty}\binom{2m}{m}
  \left(-\frac{\sigma}{2}\right)^m-1
  = \frac{1}{\sqrt{1+2\sigma}} - 1.
\end{align}
The final
equality follows from the central-binomial relation 
$\sum_{m=0}^{\infty}\binom{2m}{m}y^m=(1-4y)^{-1/2}$ with
$y=-\sigma/2$.

The Borel transform $\mathcal{B}[F](\sigma)$ has a
unique finite singularity, a square-root branch point at
$\sigma_\star=-1/2$ on the negative real axis.  
Hence, the Laplace contour
along the nonnegative real axis is entirely unobstructed.
The original formal series can therefore be recovered asymptotically from its
Borel transform by a directional Laplace integral.  Denoting the resulting
Borel-resummed function again by $F$,
\begin{equation}\label{eq:Borel_sum_F}
  F(x) := \mathcal{S}_{\mathrm B}F(x)
  = \frac{1}{x}\int_0^{\infty}
  e^{-\sigma/x}\,\mathcal{B}[F](\sigma)\,d\sigma.
\end{equation}
After the change of variables $\sigma=xt$, expanding
$\mathcal{B}[F](xt)$ and using
$\int_0^\infty e^{-t}t^m\,dt=m!$ reproduces $F(x)$ term by term as
$x\to0^+$.  Thus $F$ is \emph{strictly Borel summable} along the positive
real direction, with no Stokes phenomenon in that direction and no ambiguity
requiring any lateral deformation of the contour.

\subsection{Closed-form Borel sum and dispersion relation}
Substituting Eq.~\eqref{eq:Borel_F} into Eq.~\eqref{eq:Borel_sum_F} reduces the Laplace integral to a
Gaussian integral, yielding
\begin{equation}\label{eq:SBF_closed}
  \mathcal{S}_{\mathrm B}F(x)
  = \sqrt{\tfrac{\pi}{2x}}\;
  \mathrm{erfcx}\!\left(\tfrac{1}{\sqrt{2x}}\right) - 1,
\end{equation}
where $\mathrm{erfcx}(y):=e^{y^2}\mathrm{erfc}(y)$.  
Substituting this sum for $F$ in Eq.~\eqref{eq:self_consistency} gives the
resummed CE dispersion relation, with the hydrodynamic branch selected by
$\hat{\omega}(k)\to0$ as $k\to0$:
\begin{equation}\label{eq:omega_closed}
  \hat{\omega}+1 = \sqrt{\tfrac{\pi}{2}}\;\frac{1+\hat{\omega}}{k}\;
  \mathrm{erfcx}\!\left(\frac{1+\hat{\omega}}{k\sqrt{2}}\right).
\end{equation}
The Borel sum determines $\hat\omega(k)$ without truncating the CE
series.  To evaluate and plot this implicit relation, we introduce
$\chi=(1+\hat\omega)/k$.  Equation~\eqref{eq:omega_closed} then takes the
parametric form
\begin{equation}\label{eq:omega_parametric}
  k(\chi)=\sqrt{\frac{\pi}{2}}\,\mathrm{erfcx}\!\left(\frac{\chi}{\sqrt2}\right),
  \qquad
  \hat\omega(\chi)=\chi\,k(\chi)-1.
\end{equation}
Therefore, the dispersion relation can be traced by varying the single real
parameter $\chi$, without solving an implicit equation at each value of $k$.
At $\chi=0$, where $\hat\omega=-1$, Eq.~\eqref{eq:omega_parametric} and
$\mathrm{erfcx}(0)=1$ give
\begin{equation}\label{eq:kcrit}
  k_{\rm crit}=\sqrt{\frac{\pi}{2}}.
\end{equation}
This agrees with the critical wavenumber obtained from the spectral
Jacobi-operator analysis of
Refs.~\cite{kogelbauer2021nonlocal,kogelbauer2024rigorous,kogelbauer2025relation}.
Negative values of $\chi$ continue the branch to $\hat\omega<-1$ and
$k>k_{\rm crit}$, beyond the point where the hydrodynamic mode is an isolated
spectral point.  The physical interpretation of this continuation is left to
future work.

Figure~\ref{fig:attractor} shows the resulting dispersion relation.  The red
curve is the closed-form Borel sum: it reconstructs the hydrodynamic branch for
$0<k<k_{\rm crit}$ and continues it beyond $k_{\rm crit}$.  
The roots of the
recurrence polynomials \eqref{eq:poly}, $P_n(\hat\omega,k^2)=0$, approach this curve as $n$
increases, and their turning points move toward $k_{\rm crit}$.  In contrast,
the finite-order CE approximations $\hat\omega=-k^2$ and
$\hat\omega=-k^2+k^4$ depart from the exact branch at moderate $k$, reflecting the
known divergence of diffusion and Burnett-type expansions~\cite{Karlin2002,struchtrup2005}.

\subsection{Large-order behavior of the CE coefficients}
\label{sec:largeorder}
The Lagrange inversion~\eqref{eq:a2n_exact} can be recast in a form
better suited to large-order analysis.  Using
$F'(1{+}F)^{-2n} = -(2n{-}1)^{-1}\,\frac{d}{dx}(1{+}F)^{1-2n}$
together with $[x^{n-1}]G'(x) = n\,[x^{n}]G(x)$ gives the equivalent
representation
\begin{equation}\label{eq:a2n_alt}
  a_{2n} = -\frac{1}{2n-1}\,[x^{n}]\,\bigl(1+F(x)\bigr)^{-(2n-1)}.
\end{equation}
Expanding the power gives
\begin{equation*}
(1+F)^{-(2n-1)} = \sum_{j\ge0} (-1)^j \binom{2n+j-2}{j} F^j.
\end{equation*}
For fixed $j$ and large $n$, the binomial coefficient behaves as
\begin{equation*}
\binom{2n+j-2}{j} \sim \frac{(2n)^{j}}{j!}.
\end{equation*}
Therefore,
\begin{equation*}
\frac{a_{2n}}{(-1)^n(2n-1)!!} \sim \sum_{j\ge1}\frac{(-1)^{j+1}}{(j-1)!}=e^{-1}.
\end{equation*}
Thus the leading large-order behavior is
\begin{equation*}
a_{2n} \sim (-1)^n (2n-1)!!/e.
\end{equation*}
Consequently, the ratio of successive coefficient magnitudes satisfies
\begin{equation}
r_n:=\left|\frac{a_{2(n+1)}}{a_{2n}}\right|\sim2n+1.
\end{equation}
In Fig.~\ref{fig:ce_coeffs}, we show the ratio $r_n$, confirming the asymptotic behavior derived above.
Thus the CE series has zero radius of convergence, while its negative-axis
Borel singularity makes it Borel summable.

\subsection{Contrast with temporal gradient expansions}

The Borel summability found here is a \textit{model-specific}
feature of the linear one-dimensional BGK system and
contrasts with the temporal expansion for Bjorken flow in M\"uller--Israel--Stewart theory.
In that model, the leading Borel singularity lies on the positive real axis at
$\xi_0=3/(2C_{\tau\Pi})>0$,
where it is associated with the decay of a non-hydrodynamic mode~\cite{heller2015hydrodynamics}. 
The positive-axis singularity obstructs the Laplace contour, rendering the temporal series non-Borel summable in that direction and requiring a transseries completion.

The spatial problem considered here has a different source of divergence: factorial growth arises from the unbounded velocity tail of the Gaussian equilibrium rather than from a macroscopic non-hydrodynamic mode.
The geometric expansion of the velocity-space in Eq.~\eqref{eq:self_consistency}
produces the alternating CE coefficients and places the leading Borel singularity on the negative real axis, leaving the positive Laplace contour unobstructed. Consequently, within this model, the short-wavelength instabilities of certain higher-order CE truncations~\cite{bobylev2006instabilities} are truncation artifacts of an alternating, divergent but Borel-summable series, while the resummed relation remains well defined.

\section{Spatial Gradient Expansion for Compact Velocity Support}
\label{sec:bounded}
Since the divergence of the spatial gradient expansion is tied to the Gaussian velocity distribution, it is not surprising that restricting the velocity support to a compact domain can lead to a series convergent near $k=0$.

Consider the bounded-velocity analogue of the BGK model~\eqref{eq:BGK_nondim}:
\begin{equation}\label{eq:bounded_BGK}
  \partial_t f = -v\partial_x f  -(f - n\,W(v)),
\end{equation}
where $W(v)$ is a normalized, symmetric, nonnegative equilibrium weight supported on a compact interval $v \in [-v_{\max}, v_{\max}]$. For convenience, we set $v_{\max}=1$. The locally conserved particle density is $n(x,t) = \int_{-1}^1 f(x,v,t)\,dv$.
The motivation for restricting the velocity support to $|v| \le 1$ comes from relativistic causality, which enforces a strict speed-of-light bound. A standard relaxation-time approximation (RTA) in
relativistic kinetic theory is provided by the Anderson--Witting model~\cite{anderson1974relativistic}. We emphasize, however, that matching only the particle density $n(x,t)$ in Eq.~\eqref{eq:bounded_BGK} does
not conserve energy and momentum. Equation~\eqref{eq:bounded_BGK} is therefore not a complete relativistic kinetic theory, but rather a scalar, bounded-velocity relaxation model designed to isolate the
mathematical role of compact velocity support.

The equilibrium velocity moments of $W(v)$ are defined as
\begin{equation}
      \mu_l = \int_{-1}^{1} v^l W(v)\,dv, \qquad l = 0,1,2,\dots
\end{equation}
By symmetry $\mu_{2m+1}=0$, normalization imposes $\mu_0=1$, and compact support on $[-1,1]$ guarantees that $|\mu_{2m}| \le 1$ for all $m \ge 0$.

The evolution of the generating function $\hat{Z}(\lambda, k, t) = \int_{-1}^{1} e^{-i\lambda v}\hat{f}(k,v,t)\,dv$ takes a similar form to Eqs.~\eqref{eq:Zhat_plus}--\eqref{eq:Zhat_minus} with one change: the Gaussian factor $e^{-\lambda^2/2}$ is replaced by the
$\hat{W}(\lambda) = \int_{-1}^{1} e^{-i\lambda v} W(v)\,dv$.
Imposing the dynamic invariance condition on the Fourier-transformed kinetic equation, solving for the invariant distribution $\hat{f}$, and integrating over $v \in [-1, 1]$ yields the exact self-consistency condition
\begin{equation}\label{eq:self_consistency_bounded}
  1 = \int_{-1}^{1} \frac{W(v)}{1 + \hat{\omega} + ikv}\,dv
\end{equation}
Expanding the resolvent in powers of $x = k^2(1{+}\hat{\omega})^{-2}$ yields the auxiliary implicit relation,
\begin{equation}\label{eq:F_supp}
    \hat{\omega} =  \sum_{m=1}^{\infty} (-1)^m \mu_{2m}\, x^m:=F_{\rm supp}(x) 
\end{equation}
Equation~\eqref{eq:F_supp} is the compact velocity support version of Eq.~\eqref{eq:self_consistency}, and CE coefficients follow from the same Lagrange inversion~\eqref{eq:a2n_exact} replacing $F$ with $F_{\rm supp}$.
The crucial difference is that the bounded support $v \in [-1,1]$ guarantees $\mu_{2m} \leq 1$ for all $m$, in contrast to the factorially growing Gaussian moments $(2m{-}1)!!$.
Consequently, $F_{\rm supp}(x)$ has a strictly non-zero radius of convergence, and by the inverse function theorem the CE series $\hat{\omega}(k) = \sum a_{2n} k^{2n}$ also converges.
The spatial gradient expansion is therefore convergent in a neighborhood of $k=0$.

\section{Conclusion and Outlook}
\label{sec:conclusion}
The linear one-dimensional BGK model is simple enough that its spatial
gradient expansion can be followed to all orders and summed in closed form.
The Chapman--Enskog series diverges factorially, yet we show the Borel transform is
singular only on the negative real axis, so the Laplace integral along the
positive direction reconstructs the hydrodynamic dispersion branch without
ambiguity.
Temporal gradient expansions obstructed on the positive axis, and spatial
expansions with a finite radius of convergence, are both documented in the
literature.  In this work, we present a third possibility for a spatial
gradient expansion: factorially divergent, yet Borel summable in the
physical direction.
Because the Borel sum is available in closed form, the dispersion relation
is obtained explicitly rather than numerically. It agrees with the spectral
construction wherever the hydrodynamic mode is an isolated spectral point.
Beyond the critical wavenumber, it provides an analytic continuation of the
dispersion branch that no longer corresponds to an isolated hydrodynamic mode.

The divergence has a kinematic origin: it stems from the factorially growing
velocity moments of the Maxwellian equilibrium rather than from any
non-hydrodynamic mode.  
Bounding the velocity support
removes this growth at its source and returns a
series convergent near $k=0$.

Whether the Borel summability with negative singularity survives once energy and momentum are
restored as conserved fields, and several hydrodynamic branches couple, is the
immediate question.  A relaxation-time approximation respecting the
microscopic conservation laws~\cite{rocha2021novel}, and a full Boltzmann
collision integral would test it further.

\section*{Acknowledgements}
Author thanks C.~E.~Frouzakis for comments on the initial draft and
I.~V.~Karlin and F.~Kogelbauer, for insightful correspondence and for drawing attention to the
relevant spectral results.
\bibliography{references}
\end{document}